\documentclass[twocolumn,showpacs,amsmath,amssymb,superscriptaddress]{revtex4-2}
\usepackage{longtable} 
\usepackage[dvips]{graphicx}
\usepackage{amssymb,amsfonts,amsmath}
\usepackage{comment}
\usepackage{here}
\usepackage{xcolor}
\usepackage[normalem]{ulem}

\begin{document}%
\title{Power-law molecular-weight distributions dictate universal behaviors in highly polydisperse polymer solutions}

\author{Naoya Yanagisawa$^*$}
\affiliation{Komaba Institute for Science, Graduate School of Arts and Sciences, The University of Tokyo, Komaba 3-8-1, Meguro, Tokyo 153-8902, Japan}

\author{Daisuke S. Shimamoto}
\affiliation{Komaba Institute for Science, Graduate School of Arts and Sciences, The University of Tokyo, Komaba 3-8-1, Meguro, Tokyo 153-8902, Japan}

\author{Miho Yanagisawa$^*$}
\affiliation{Komaba Institute for Science, Graduate School of Arts and Sciences, The University of Tokyo, Komaba 3-8-1, Meguro, Tokyo 153-8902, Japan}
\affiliation{Department of Physics, Graduate School of Science, The University of Tokyo, Hongo 7-3-1, Bunkyo, Tokyo 113-0033, Japan}
\affiliation{Center for Complex Systems Biology, Universal Biology Institute, The University of Tokyo, Komaba 3-8-1, Meguro, Tokyo 153-8902, Japan}

\date{\today}

\begin{abstract}
Polydispersity is a universal feature of synthetic polymers and biological molecules in the cytoplasm. However, its quantitative impact on collective behavior remains poorly understood because conventional metrics, such as the polydispersity index, fail to capture broad, non-Gaussian size distributions. Here, we develop an experimental platform in which polyethylene glycol (PEG) solutions are engineered to follow tunable power-law molecular-weight distributions spanning an extensive range, from $M = 1$~kg/mol to $10^{4}$~kg/mol. By systematically varying the $M$ distribution exponent $a$, we identify a robust regime ($1 < a \lesssim 2.5$) in which the viscosity scaling exponent in the entangled regime, the overlap concentration $c^{\ast}$, and the entanglement concentration ${c_{\mathrm{e}}}$ all exhibit pronounced maxima that exceed monodisperse limits. This amplification minimizes as the upper cutoff $M_{\max}$ is reduced, with the system approaching monodisperse behavior. The enhanced rheology arises from a competition between long-chain-dominated entanglement and short-chain-mediated void filling, demonstrating that the whole shape of the molecular-weight distribution plays a decisive role. 
Consequently, these collective behaviors cannot be reproduced by simply tuning the average molecular weight. Together, our results establish the power-law exponent $a$ as a quantitative control parameter that links polymer entanglement, soft packing, and molecular crowding in highly polydisperse systems.
\end{abstract}

\maketitle

\section*{Introduction}
Natural and synthetic materials rarely consist of identical units; instead, they contain components spanning broad size or molecular-weight distributions. This polydispersity is intrinsic to systems ranging from colloidal suspensions~\cite{sollich2001predicting,royall2024colloidal} to biological molecules in living cells, whose molecular weights are broadly distributed over several orders of magnitude from small amino acids, $\sim 0.1\ \mathrm{kDa}$, to large proteins and complexes, $\sim 10^{3}\ \mathrm{kDa}$~\cite{uversky2017intrinsically,hatton2023human}.
Despite extensive research on rheological properties of such polydisperse solutions~\cite{powell1966influence,cross1969polymer,nichetti1998viscosity,cassagnau1993rheology,stadler2006dependence,vega2012effect,blanco2018polydispersity} and years of highlighting the importance of polydispersity~\cite{dunleavy1966correlation,cross1969polymer,wasserman1992effects,peters2018effect,alfano2024molecular}, it has often been regarded as an experimental complication rather than recognized as a fundamental physical variable that determines the organization and biochemical functions of the system.

In particulate systems, size polydispersity allows smaller particles to occupy interstitial voids between larger ones, which increases packing efficiency and stabilizes disordered states through a “void-filling’’ mechanism~\cite{torquato2010jammed,berthier2016equilibrium,ninarello2017models}. As a result, polydisperse assemblies often achieve higher packing fractions at higher jamming thresholds than monodisperse systems~\cite{oquendo2020densest,shimamoto2023common}. These observations suggest that polydispersity can fundamentally reshape collective behavior. Similar principles are expected to operate in polymeric and biological environments, where molecular-weight distributions naturally emerge from synthesis, degradation, and nonequilibrium intracellular processes. Cell extracts exhibit broad, non-Gaussian molecular-weight distributions and display anomalous viscoelasticity, crowding-induced segregation, and unconventional phase behavior~\cite{nishizawa2017universal,zhou2008macromolecular,marenduzzo2006depletion,fabry2001scaling,wilhelm2008out,huang2022cytoplasmic}. However, a systematic framework for controlling and quantifying the effects of continuous molecular-weight polydispersity on polymer rheology has remained elusive.

Recent advances have featured the fundamental limitations in describing polydispersity using parameters that evaluate deviation from an average, such as the variance or the polydispersity index (PDI = $M_{\mathrm w}/M_{\mathrm n}$, where $M_{\mathrm w}$ and $M_{\mathrm n}$ denote the weight- and number-average molecular weights, respectively)~\cite{rogovsic1996polydispersity,gentekos2019controlling,rane2005polydispersity,harrisson2018downside,gentekos2016beyond}. These descriptors are effective when the molecular weight distributions are narrowly concentrated around a characteristic size~\cite{rane2005polydispersity,harrisson2018downside}. However, they fail to capture the structure of broad non-Gaussian distributions that arise in highly polydisperse systems without a characteristic size.

A more general framework is provided by continuous distributions, particularly power-law forms $N(M)\propto M^{-a}$, characterized by an exponent $a$. For sufficiently broad power-law distributions, statistical moments such as the mean and variance can become ill-defined~\cite{newman2005power}. For the $M$ distribution, $M_{\mathrm n}$ diverges for $a\le2$ and $M_{\mathrm w}$ diverges for $a\le3$, indicating that minor populations can dominate macroscopic behavior~\cite{clauset2009power}. To ensure a finite total number of chains, $a$ must exceed unity. This perspective has been validated in particulate systems, where specific ranges of $a$ yield denser packings and higher jamming points than monodisperse assemblies~\cite{shimamoto2023common,shimamoto2024preparation,torquato2010jammed,oquendo2020densest}. These findings suggest that the exponent $a$, rather than average-based metrics such as PDI, serves as a natural control parameter for collective behavior in highly polydisperse matter~\cite{sollich2001predicting}.

Building on this insight, we focus on highly polydisperse polymer solutions, specifically those lacking a clear characteristic molecular-weight scale. Power-law distributions with fixed lower and upper cutoffs ($M_\mathrm{min}$ and $M_\mathrm{max}$) generate a series of molecular-weight distributions parametrized by the exponent $a$, smoothly connecting the two monodisperse limits dominated by either $M_\mathrm{min}$ or $M_\mathrm{max}$. While the polydisperse system approaches effective monodispersity near these limits, the notion of a typical molecular weight becomes ambiguous in the intermediate $a$ regime, which is expected to satisfy the divergence condition for $M_{\mathrm n}$ ($a\le2$) and $M_{\mathrm w}$ ($a\le3$)~\cite{newman2005power}. This framework allows us to systematically place polydisperse polymer systems on the path, rather than outside, of existing theoretical understanding of polymers characterized by $M_{\mathrm w}$.

Here, we develop an experimental platform to explore this regime directly using polyethylene glycol (PEG) solutions with tunable power-law molecular-weight distributions spanning several orders of magnitude under fixed cutoffs. By systematically varying the exponent, $a$, we probe how rheological properties emerge in polymer solutions without a well-defined molecular-weight scale, focusing on the onset of overlap, entanglement, and macroscopic flow behavior. This approach allows us to isolate the role of distribution shape, independent of average molecular weight, in governing polymer rheology.

More broadly, this study positions the exponent $a$ as a quantitative parameter that links polymer crowding and entanglement to concepts developed initially in idealized model systems, including athermal particles. By framing molecular-weight polydispersity as a tunable physical variable rather than a perturbation around an average, our results provide a unified perspective on multi-component mixtures, including the crowded intracellular environment.

\begin{figure*}[t]
\centering
\includegraphics[width=\linewidth]{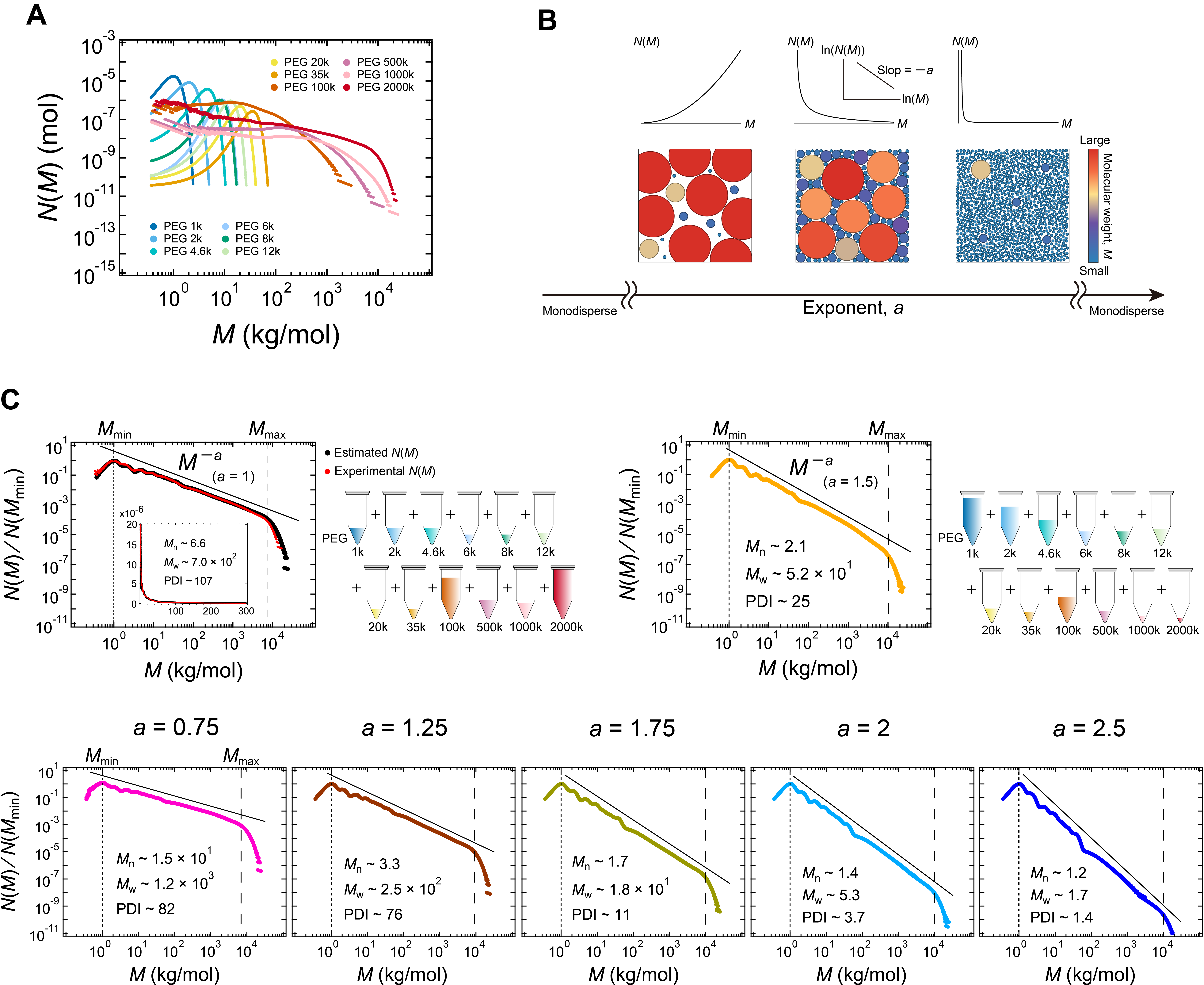}
\caption{
Preparation and characterization of PEG samples with power-law molecular-weight distributions. 
(A) Molecular-weight, $M$, number distributions of twelve PEG samples ranging from 1k to 2000k (see S2 for details). 
(B) Schematic illustrating how the molecular-weight exponent $a$ shapes polydispersity. 
Small $a$ enriches large-$M$ polymers, whereas large $a$ enriches small-$M$ polymers. Both limits effectively approaching monodispersity. 
At intermediate $a$, neither population dominates, producing genuinely broad and enhanced polydispersity. 
(C) (Top left) Example of a power-law $M$ distribution with $a = 1$, prepared by blending the twelve PEG samples in specified mass ratios (Fig.~S2). 
The experimentally measured distribution (red) agrees well with the computational prediction (black). 
The inset shows the same distribution on a linear scale. 
Number-average molecular weight $M_{\mathrm n}$, weight-average molecular weight $M_{\mathrm w}$, and the polydispersity index (PDI = $M_{\mathrm w}/M_{\mathrm n}$) are indicated. 
(Top right and bottom) Normalized $M$ distributions for six polydisperse PEG samples with different $a$, all sharing identical cutoff values ($M_{\mathrm{min}} = 1$~kg/mol, $M_{\mathrm{max}} \sim 1 \times 10^4$~kg/mol), shown by the dotted and dashed lines.
}
\label{Fig1}
\end{figure*}

\begin{figure*}[t]
\centering
\includegraphics[width=\linewidth]{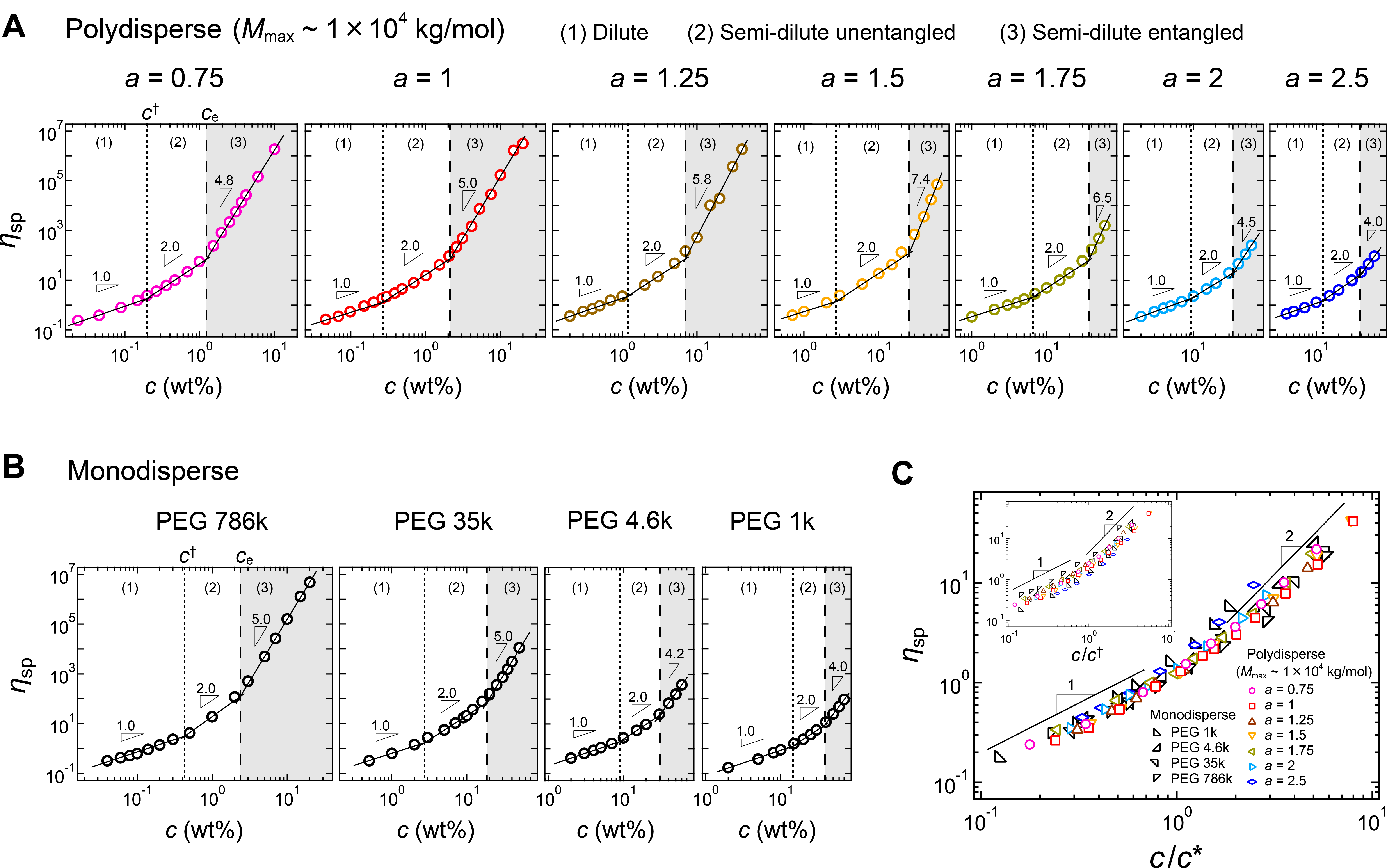}
\caption{
Concentration-dependent specific viscosity, $\eta_{\mathrm{sp}}$, of polydisperse and monodisperse PEG solutions.
(A) $\eta_{\mathrm{sp}}$ as a function of polymer concentration $c$ for polydisperse PEG samples with a fixed upper cutoff $M_{\mathrm{max}} \sim 1 \times 10^4$~kg/mol and different molecular-weight exponents $a$ (see Fig.~S6 for polydisperse samples with other values of $M_{\mathrm{max}}$). 
(B) Corresponding data for monodisperse PEG samples with molecular weights $M$ of 786k, 35k, 4.6k and 1k (PDI $< 1.3$) (see Fig.~S5 for monodisperse samples with $M$ of 12k, 8k, 6k, and 2k).
(C) $\eta_{\mathrm{sp}}$ plotted against the scaled concentrations $c/c^{\ast}$, where $c^{\ast}$ is the overlap concentration; the inset shows the same data plotted against $c/c^{\dagger}$, where $c^{\dagger}$ marks the crossover between the dilute and semi-dilute regimes.
Across all samples, $\eta_{\mathrm{sp}}$ exhibits three well-established scaling regimes under $\Theta$-solvent conditions (black lines): 
(1) $c^{1}$ for $c < c^{\ast}$ (dilute), 
(2) $c^{2}$ for $c^{\ast} < c < c_{\mathrm{e}}$ (semi-dilute unentangled), and 
(3) $c^{\simeq 5}$ (or $c^{\simeq 4}$ for short polymers with $M_\mathrm{w} < 5$~kg/mol) for $c > c_{\mathrm{e}}$ (semi-dilute entangled). 
In panels (A) and (B), dotted and dashed vertical lines indicate the crossover concentrations $c^{\dagger}$ and $c_{\mathrm{e}}$, separating regimes (1)–(2) and (2)–(3), respectively.
Shaded regions highlight the entangled regime (3), where the viscosity scaling becomes strongly dependent on $a$. 
}
\label{Fig2}
\end{figure*}

\section*{Materials and Methods}
\subsection*{Materials}
Polyethylene glycol (PEG; also known as polyethylene oxide, PEO) samples with different molecular weights $M$ were obtained from commercial suppliers (FUJIFILM Wako Pure Chemical Corporation, Osaka, Japan, unless otherwise noted). The samples included PEG 1k, 2k, 4.6k, 6k, 8k, 12k, 20k, 35k, PEO 100k, PEG 500k, PEO 1000k, and PEG 2000k (see S1 in the Supporting Information, SI). For relatively short PEGs ($M\leq 35$ kg/mol), the $M$ distribution was assumed to be Gaussian and centered at the nominal $M$ (see S2). The PDI of these short PEG samples was below 1.3. For long PEGs ($M\geq 100$ kg/mol), the $M$ distributions were independently characterized using gel permeation chromatography (GPC), with representative distributions shown in Fig.~S1. For monodisperse samples, PEG 786k (PDI = 1.16; analytical standards for GPC; Tosoh Co., Japan) was used.

\subsection*{Polymer solution preparation}
To prepare polydisperse PEG solutions with controlled power-law molecular-weight distributions, selected PEG samples were mixed in specified mass ratios (see S2 in SI). By adjusting the mixing ratios, we systematically tuned the power-law exponent $a$ over the range $0.75 \leq a \leq 2.5$, which satisfies the divergence condition $a \leq 3$ of the average molecular weight in an ideal system without cutoffs~\cite{newman2005power}.
Extending $a$ beyond this range was experimentally impractical because the contribution of the minor component becomes negligibly small. Throughout this study, the minimum molecular weight was fixed at $M_{\mathrm{min}} = 1$~kg/mol. The maximum molecular weight was set to $M_{\mathrm{max}} \sim 1 \times 10^{4}$~kg/mol for most samples, and was reduced to $\sim 1 \times 10^{3}$~kg/mol, $\sim 2 \times 10^{2}$~kg/mol, or $\sim$ 45~kg/mol in selected cases to systematically narrow the distribution width. The resulting number-average molecular weight $M_\mathrm{n}$, weight-average molecular weight $M_\mathrm{w}$, and polydispersity index (PDI) for each sample are summarized in Fig.~S1 and S3.
All PEG samples were dissolved in ultrapure distilled water (Invitrogen, Waltham, MA, USA) to prepare aqueous polymer solutions. The mixtures were stirred using a vortex mixer (HS120318, Heathrow Scientific Co.) for at least 24~h to ensure complete dissolution. PEG solutions were prepared at concentrations of up to approximately 60~wt\%, below which no crystallisation was observed. Solution densities were measured using a density meter (DMA~1001, Anton Paar GmbH) and found to depend solely on the PEG concentration, showing negligible dependence on $a$ or $M_{\mathrm{max}}$. Typical density values ranged from 1.0 to 1.1~g/cm$^{3}$.

\subsection*{Rheological measurements}
Rheological measurements were performed at $25\,^{\circ}\mathrm{C}$ using cone-plate rotational rheometers (MCR~502 and ViscoQC~300~L; Anton Paar GmbH, Graz, Austria). Both steady-shear and oscillatory shear experiments were conducted. For oscillatory measurements, the strain amplitude was fixed at 0.5\%. The shear-rate-dependent viscosity $\eta(\dot{\gamma})$ and the frequency-dependent complex viscosity $\eta^{*}(\omega)$ were recorded. Representative flow curves are shown in Fig.~S4. At sufficiently low shear rates (or low angular frequencies), all samples exhibited a clear Newtonian plateau characterized by a constant viscosity~\cite{ebagninin2009rheological,holyst2009scaling,wisniewska2014scaling,wisniewska2017scaling}.
In this study, we focus on this Newtonian regime and analyze the dependence of the zero-shear viscosity $\eta_{0}$ on molecular-weight polydispersity. The effects of the power-law exponent $a$ and the molecular-weight distribution width on $\eta_{0}$ were systematically examined across different concentration regimes.

\section*{Results}
\subsection*{Impact of $M$ polydispersity on solution viscosity}

To investigate how high molecular-weight polydispersity influences the rheology of polymer solutions, we constructed an experimental platform based on polyethylene glycol (PEG) ensembles with tunable power-law molecular-weight distributions. Twelve commercially available PEGs (1k-2000k) spanning four orders of magnitude in molecular weight were used as building blocks (Fig.~\ref{Fig1}A). By blending their aqueous solutions in prescribed mass ratios (see Fig.S2), we generated PEG mixtures whose molecular-weight distributions follow power laws with exponent $a$. Each mixture shares the same lower cutoff ($M_{\mathrm{min}} = 1$~kg/mol) and an upper cutoff $M_{\mathrm{max}} \sim 1 \times 10^{4}$~kg/mol. These cutoff values reflect the limits imposed by available PEG species and are approximate.

Figure~\ref{Fig1}B illustrates how the molecular-weight exponent $a$ shapes polydispersity. When $a$ is small, large-$M$ polymers (close to $M_{\mathrm{max}}$) dominate the distribution; when $a$ is large, small-$M$ polymers (close to $M_{\mathrm{min}}$) dominate. Both limits effectively approaching monodispersity. At intermediate $a$ values, the mean and variance diverge, indicating that neither population dominates and producing genuinely broad, enhanced polydispersity. In an ideal system with no cutoff, the average molecular weight diverges for $a \leq 3$~\cite{newman2005power}.

Figure~\ref{Fig1}C (top left) shows a representative of experimentally obtained distribution for $a = 1$, where the measured profile (red) closely matches the computational prediction (black), confirming reproducibility of the designed polydisperse ensembles. Figure~\ref{Fig1}C (top right and bottom) summarizes six distributions with varying $a$ (0.75–2.5), all sharing the cutoff values (i.e., $M_{\mathrm{min}}$ and $M_{\mathrm{max}}$). Distributions with smaller $M_{\mathrm{max}}$ are provided in Fig.~S3.
As expected from the mathematical form of a power law, the conventional polydispersity index (PDI = $M_\mathrm{w}/M_\mathrm{n}$) decreases monotonically with increasing $a$. However, despite the decrease in PDI, the molecular weight distribution becomes significantly long-tailed, indicating a possible increase in functional polydispersity.

To assess how the exponent $a$ governs a polymer's macroscopic behavior, we measured the concentration dependence of the zero-shear viscosity $\eta_0$ for such samples. Although broad $M$ distributions induce shear thinning (Fig.~S4), here we focus on the specific viscosity 
$\eta_{\mathrm{sp}} = (\eta_0 - \eta_{\mathrm{w}})/\eta_{\mathrm{w}}$ (where $\eta_{\mathrm{w}}$ is water viscosity)
as a function of concentration $c$. Figure~\ref{Fig2}A compares $\eta_{\mathrm{sp}}(c)$ across polydisperse samples with different $a$ but identical cutoff values. For reference, Fig.~\ref{Fig2}B shows monodisperse PEGs (PDI $<1.3$) with distinct molecular weights (786k, 35k, 4.6k and 1k). Monodisperse samples with $M$ of 12k, 8k, 6k and 2k are shown in Fig.~S5

For neutral polymer solutions under $\Theta$-solvent conditions, such as the PEG solutions used here, the concentration dependence of the specific viscosity $\eta_{\mathrm{sp}}$ is known to fall into three regimes~\cite{indei2022microrheological,colby2023specific,colby2010structure}:
\[
\eta_{\mathrm{sp}} \sim
\begin{cases}
c^{1}, & c < c^\ast \quad \text{(dilute)}, \\
c^{2}, & c^\ast < c < c_{\mathrm{e}} \quad \text{(semi-dilute unentangled)}, \\
c^{\simeq 5\ (\text{or }4)}, & c > c_{\mathrm{e}} \quad \text{(semi-dilute entangled)},
\end{cases}
\]
where $c^\ast$ and $c_{\mathrm{e}}$ denote the overlap and entanglement concentrations, respectively.  
The entanglement molecular weight of PEG in the molten state is known to be $\sim 2$~kg/mol and is expected to increase to $\sim 5$~kg/mol in solutions ($>40$~wt\%)~\cite{fetters1994connection,ianniello2022linear}. Consequently, for short PEG ($M_{\mathrm{w}} < 5$~kg/mol), the exponent in regime (3) decreases to approximately 4 due to the absence of entanglements, consistent with our observation that these samples show no entangled behavior (see Fig.~\ref{Fig2}B and Fig.~S5).

Monodisperse PEG samples follow these expected scaling behaviors under $\Theta$-solvent conditions: the exponent increases from 1 to 2 and then to approximately 5 (or 4 for short PEG) as concentration increases (Fig.~\ref{Fig2}B), in agreement with previous studies~\cite{indei2022microrheological}.  
In contrast, polydisperse PEG samples with different $a$ values exhibit a distinct trend (see Fig.~S6 for different $M_{\mathrm{max}}$). As shown in Fig.~\ref{Fig2}A, the scaling exponents in the dilute and unentangled regimes (1 and 2) remain essentially invariant with respect to $a$, indicating that $M$ polydispersity does not alter viscosity scaling at low and intermediate concentrations.  
However, in the entangled regime (3), the scaling exponent becomes strongly dependent on the molecular-weight exponent $a$, demonstrating that continuous $M$ polydispersity qualitatively reshapes viscosity scaling only when polymers are sufficiently concentrated to form entanglements.

\begin{figure*}[t]
\centering
\includegraphics[width=\linewidth]{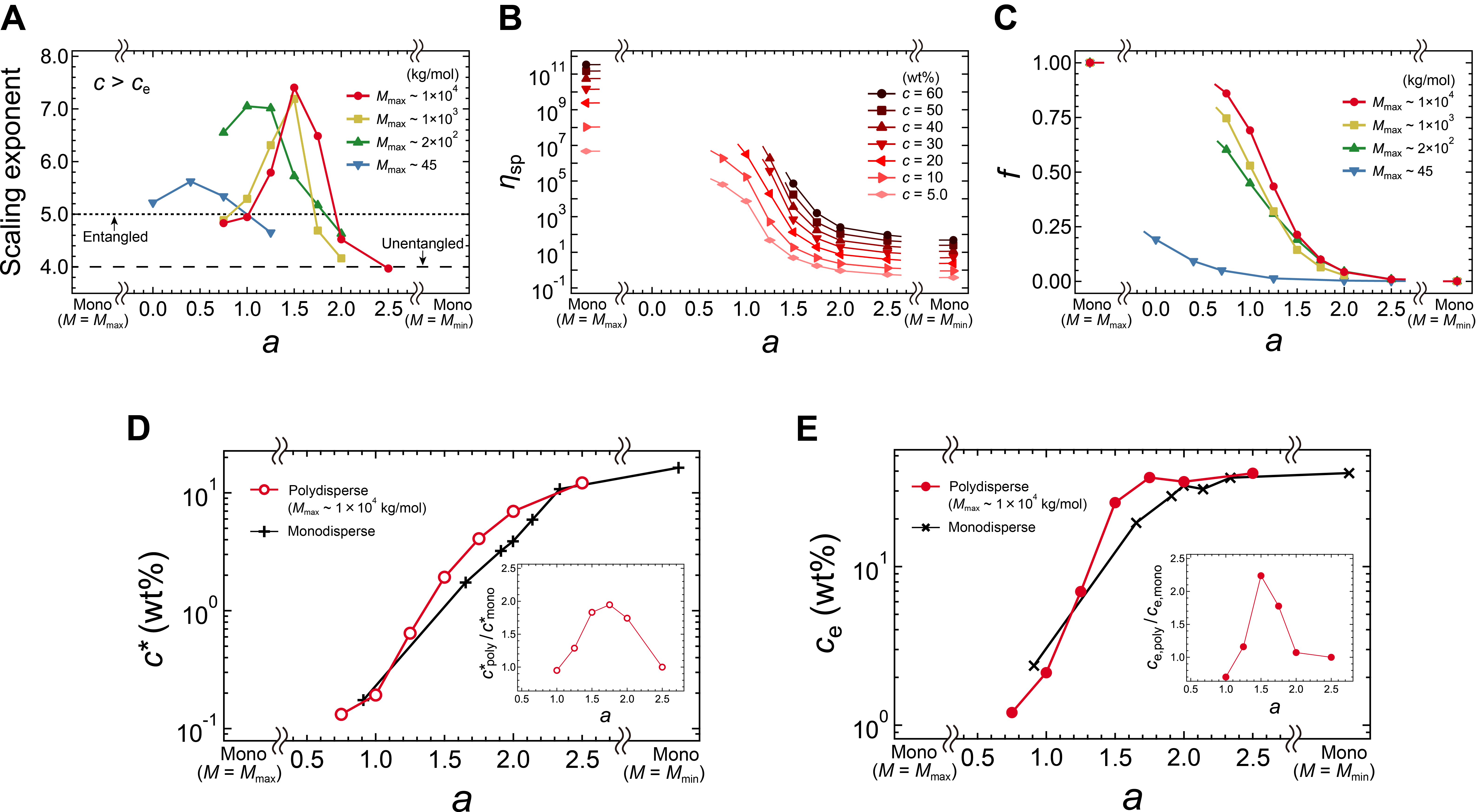}
\caption{Impact of $M$ polydispersity on the overlapping and entanglement in the semi-dilute regimes (2, 3).
    (A, B) Dependence of the viscosity scaling exponent in the regime (3) above $c_{\mathrm{e}}$ and $\eta_{\mathrm{sp}}$ on $M$ exponent, $a$. In panel (B), monodisperse $\eta_{\mathrm{sp}}$ at $M = M_{\mathrm{max}}$ is estimated from the $c^{5}$ scaling.
    (C) Relationship between the mass proportion of long-polymer components ($M > 20$~kg/mol) capable of forming entanglements $f$ and $a$. The color in (A, C) represents $M_{\mathrm{max}}$: $\sim 1 \times 10^{4}$~kg/mol (red), $\sim 1 \times 10^{3}$~kg/mol (yellow), $\sim 2 \times 10^{2}$~kg/mol (green), and $\sim 45$~kg/mol (blue). $M_{\mathrm{max}}$ for (B) is $\sim 1 \times 10^{4}$~kg/mol.
    (D, E) Dependence of $c^{*}$ and $c_{\mathrm{e}}$ on $a$ for $M_{\mathrm{max}} \sim 1 \times 10^{4}$~kg/mol. Black cross symbols denote their values for monodisperse samples with almost the same $M_\mathrm{w}$ as the corresponding polydisperse samples. The inset shows the normalized value of $c_{\mathrm{poly}}/c_{\mathrm{mono}}$, where $c_{\mathrm{mono}}$ is estimated from the monodisperse curves obtained by connecting the data points, using linear interpolation at the corresponding $a$ (see Fig.~S8 for different $M_{\mathrm{max}}$).
}
\label{Fig3}
\end{figure*}

To examine more closely how $M$ polydispersity influences viscosity scaling, we expanded the specific viscosity in a power series around $c=0$,
\[
\frac{\eta_{\mathrm{sp}}}{c} = [\eta] + k_{\mathrm{H}}[\eta]^2 c + \mathcal{O}(c^2),
\]
where $[\eta]$ and $k_{\mathrm{H}}$ denote the intrinsic viscosity and Huggins constant, respectively~\cite{RubinsteinColby2003}. The inverse intrinsic viscosity $1/[\eta]$ provides an estimate of the overlap concentration $c^\ast$, the point at which polymer coils begin to geometrically overlap~\cite{RubinsteinColby2003}.
Figure~\ref{Fig2}C shows $\eta_{\mathrm{sp}}$ as a function of the scaled concentration $c[\eta]=c/c^\ast$ for both monodisperse and polydisperse PEG samples. In the dilute and semi-dilute unentangled regimes (1 and 2), all datasets collapse onto a single master curve with slopes of 1 and 2, respectively, consistent with established scaling laws~\cite{colby2023specific,colby2010structure}. A similar collapse is obtained when the data are plotted against $c/c^{\dagger}$ (inset), indicating that the transition concentration between regimes (1) and (2), $c^{\dagger}$, is effectively equivalent to the overlap concentration $c^{\ast}$($=1/[\eta]$) in the dilute limit. Given this equivalence, we hereafter use $c^{\ast}$ to represent the transition concentration between (1) the dilute and (2) the unentangled regimes.

\subsection*{$M$ polydispersity magnified in a specific $a$ range}

To quantify the impact of high molecular-weight polydispersity on polymer entanglement, we extracted the viscosity scaling exponent in the entangled regime (regime 3) for all samples. As shown in Fig.~\ref{Fig3}A, polydisperse PEG samples with $M_{\mathrm{max}} \sim 1 \times 10^{4}$~kg/mol (red circles) exhibit a pronounced peak in the scaling exponent within $1 < a \lesssim 2.5$, exceeding the ideal values of 4–5 observed for monodisperse solutions with or without entanglement (dotted and dashed horizontal lines). The position of this maximum depends sensitively on $M_{\mathrm{max}}$. When we systematically reduced the $M_{\mathrm{max}}$ to $\sim 1 \times 10^3$~kg/mol (yellow square), $\sim 2 \times 10^2$~kg/mol (green triangle), and  $\sim 45~$kg/mol (blue inverted triangle), this peak shifts toward smaller $a$ and its width becomes broader.
As $M_{\mathrm{max}}$ approaches $M_{\mathrm{min}}$, the $M$ distribution narrows and the system becomes effectively monodisperse. This may explain why, as $M_{\mathrm{max}}$ decreases, the peaks become less distinct and the values tend to approach those of homogeneous systems (i.e., 4--5).

To explore the correspondence between the viscosity scaling exponent and the zero-shear specific viscosity $\eta_{\mathrm{sp}}$, we plotted $\eta_{\mathrm{sp}}$ against $a$ (Fig.~\ref{Fig3}B) for various concentrations, $c$, shown in different colors. Interestingly, $a$ $\sim2.5$, where $\eta_{\mathrm{sp}}$ starts to increase, corresponds to the upper boundary of a specific $a$ region where the viscosity scaling exponent becomes larger than the monodisperse value (Fig.~\ref{Fig3}A).

Regarding this agreement, we next analyzed how the exponent $a$ controls the population of long polymers that dominantly contribute to the viscosity increase. To this end, we evaluated the mass fraction $f$ of polymers with molecular weight greater than 20~kg/mol, approximately 10 times the entanglement molecular weight of molten PEG~\cite{fetters1994connection,ianniello2022linear}. As shown in Fig.~\ref{Fig3}C, for systems with $M_{\mathrm{max}} \sim 1 \times 10^{4}$~kg/mol (red symbols), the fraction $f$ increases sharply when $a \lesssim 2.5$. As $M_{\mathrm{max}}$ decreases, the onset of this increase shifts systematically toward smaller values of $a$, reflecting the truncation of the high-molecular-weight tail in the distribution (see other colors).
We confirmed that this trend remains unchanged when the threshold defining long polymers is increased (e.g., $M > 100$~kg/mol; see Fig.~S7).
Notably, both the onset values of $a$ and their systematic shift with decreasing $M_{\mathrm{max}}$ closely parallel the corresponding changes observed in the viscosity scaling exponent and the specific viscosity $\eta_{\mathrm{sp}}$ (Fig.~\ref{Fig3}(A, B)). 

Accordingly, we examined how the overlap concentration $c^{\ast}$ and the entanglement concentration $c_{\mathrm{e}}$ depend on the exponent $a$ (Fig.~\ref{Fig3}(D, E)). For polydisperse samples with $M_{\mathrm{max}} \sim 1 \times 10^{4}$~kg/mol, both $c^{\ast}$ and $c_{\mathrm{e}}$ exhibit pronounced enhancements within the range $1 < a \lesssim 2.5$, exceeding the values observed in monodisperse samples with the same $M_{\mathrm{w}}$. When normalized by the corresponding monodisperse values, both quantities display clear peaks in this specific range of $a$ (insets of Fig.~\ref{Fig3}(D, E)), mirroring the behavior of the viscosity scaling exponent (Fig.~\ref{Fig3}A).
A similar trend is observed for samples with a smaller cutoff, $M_{\mathrm{max}} \sim 1 \times 10^{3}$ and $2 \times 10^{2}$~kg/mol (see Fig.~S8), although the position and magnitude of the peaks shift systematically. This shift reflects the reduced availability of long polymers as the upper cutoff of the $M$ distribution decreases.

Taken together, these results demonstrate that three independent measures of collective polymer behavior, i.e., the viscosity scaling exponent, the overlap concentration $c^{\ast}$, and the entanglement concentration $c_{\mathrm{e}}$, reach their maximum values within a specific range of the $M$ exponent $a$ ($1 < a \lesssim 2.5$ for large $M_{\mathrm{max}}$). The location of this range is governed by the balance between long- and short-polymer populations, as reflected in the tail of the $M$ distribution. As $M_{\mathrm{max}}$ decreases and the $M$ distribution narrows, this amplification regime gradually weakens and eventually vanishes, approaching the behavior of monodisperse systems. These findings identify the exponent $a$ as a key physical parameter that determines when viscosity and entanglement are maximally amplified in highly polydisperse polymer solutions.

\section*{Discussion}

\subsection*{Heavy-tailed polydispersity controls rheology}

\begin{figure}[t]
\centering
\includegraphics[width=0.9\columnwidth]{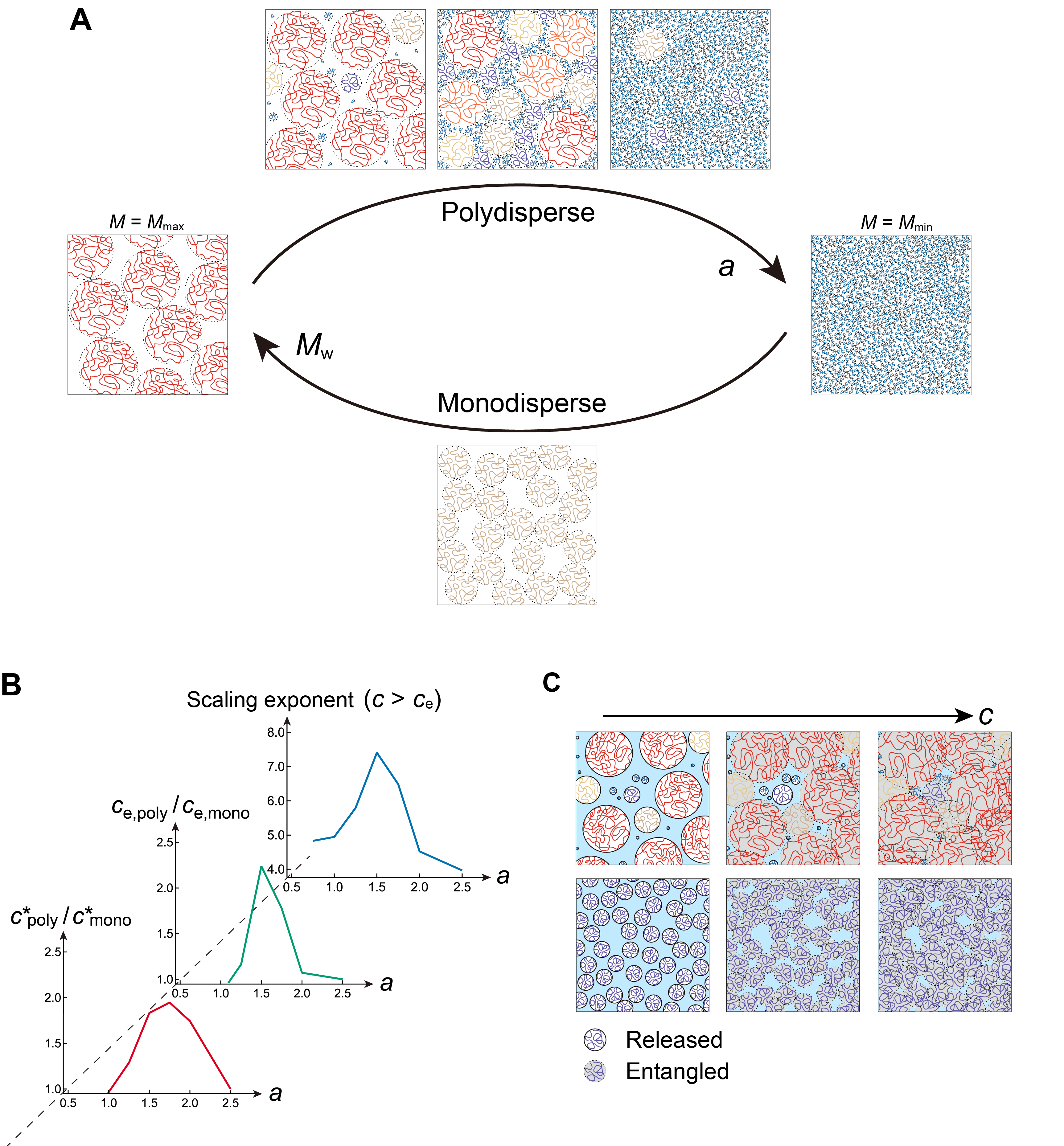}
\caption{
Conceptual schematic depicting the influence of power-law molecular-weight polydispersity on the rheological properties of polymer solutions.
(A) At large and small values of the molecular weight exponent $a$, this polydisperse system approaches monodisperse behavior determined by $M_{\mathrm{min}}$ and $M_{\mathrm{max}}$, respectively. For intermediate values of $a$, the system exhibits a characteristic behavior that cannot be achieved by varying the average $M$ alone.
(B) In polydisperse systems, key collective quantities (from bottom; the overlap concentration $c^{\ast}$, the entanglement concentration $c_{\mathrm{e}}$, and the viscosity scaling exponent) are maximally enhanced within an intermediate range of $a$ ($1 < a \lesssim 2.5$ for broad distributions with large $M_{\mathrm{max}}$) compared to monodisperse systems.
(C) In this region, increasing concentration induces sequential entanglement: long polymers near $M_{\mathrm{max}}$ first initiate entanglement, followed by shorter polymers contributing to it as the concentration increases.
}
\label{Fig4}
\end{figure}

A central outcome of this work is that the rheology of polymer solutions is governed not merely by average molecular weights or variances, but by the full shape of the $M$ distribution. In particular, power-law distributions exhibit a physically distinct regime in which statistical moments, such as the mean and variance, diverge. Within these conditions, minor components qualitatively change the rheological behavior of the solutions. Our experiments directly demonstrate that polymer solutions in this heavy-tailed regime exhibit behaviors that cannot be reproduced by tuning conventional variables such as $M_{\mathrm{w}}$ or the polydispersity index (PDI) alone (see Fig.~\ref{Fig4}A).

We identify a specific range of the molecular-weight exponent $a$ ($1 < a \lesssim 2.5$ for large $M_{\mathrm{max}}$) in which three independent quantities, the viscosity scaling exponent in the entangled regime, the overlap concentration $c^{\ast}$, and the entanglement concentration ${c_{\mathrm{e}}}$, all reach obvious maxima at $a = 1.5$ beyond monodisperse limits (Fig.~\ref{Fig4}B). 

We analyze the physical significance of the specific $a$ regime ($1 < a \lesssim 2.5$) by connecting it to the conditions under which the average description fails. This $a$ regime meets the criterion for divergence of the average molecular weight $M_{\mathrm w}$, namely $a\le3$~\cite{newman2005power}. We next consider the divergence condition for polymer size characterized by the radius of gyration, $R_{\mathrm g}$. In dilute solution, $R_{\mathrm g}$ scales with molecular weight as $R_{\mathrm g} \propto M^{\nu}$, where $\nu~\simeq 0.583$ for good solvents~\cite{wisniewska2014scaling,devanand1991asymptotic}. The molecular-weight distribution $N(M)\propto M^{-a}$ can thus be transformed into
a corresponding size distribution $N(R_{\mathrm g})\propto R_{\mathrm g}^{-b}$, where
$b = (a-1)/\nu + 1$ (see S5 for details). Applying these relationships, the specific regime for $M_{\mathrm w}$ ($1 < a \lesssim 2.5$ peaked at $a = 1.5$) translates to $R_{\mathrm g}$ as $1 < b \lesssim 3.6$, peaking at $b \simeq 1.86$. The regime below the peak at $b\simeq 1.86$ satisfies $b\le 2$, which corresponds to the divergence conditions of $M_{\mathrm n}$ and $R_{\mathrm g}$~\cite{newman2005power}. The upper and lower boundaries indicate conditions where the number and impact of long or short polymers increase, resulting in deviations from the expected behavior of the averaged systems. Notably, as $M_{\mathrm{max}}$ decreases, the peak height decreases and the upper boundary shifts to a lower $a$ (Fig.~3A). This shift in the upper boundary further indicates that the presence of sufficiently large long-chain polymers in the heavy-tailed molecular weight distribution strongly affects the upper boundary of the specific $a$ regime.

The rapid increase in the viscosity exponent (Fig.~3A) is attributed to changes in the contribution of short-chain polymers as concentration rises. At low concentrations, long-chain polymers predominantly govern entanglements, while short-chain polymers remain in solution with minimal interaction. As concentration increases, a greater proportion of short-chain polymers participate in entanglements with long-chain polymers, resulting in a marked increase in the viscosity index compared to monodisperse systems with narrow molecular weight distributions (Fig.~4C). This observation is analogous to size-polydisperse jammed particle systems, where small particles contribute to the jammed network with large particles, thereby increasing the packing fraction relative to monodisperse systems~\cite{shimamoto2024compression}. In contrast to jammed particle systems, the heterogeneity inferred from absorbance data for polydisperse solutions in the entangled regime (3) may also play a role (see S4 and Fig.~S9). One hypothesis is that short-chain polymers enhance friction between aggregated long-chain polymers, which substantially increases the viscosity index. These hypotheses could be tested through direct observation using polymers labeled with distinct fluorescent molecules corresponding to specific chain lengths.

\subsection*{Connection to jamming and universal behavior in polydisperse systems}

The amplification of rheological properties observed here has a close conceptual parallel with jamming phenomena in polydisperse particulate systems. By converting the $M$ exponent $a$ to the corresponding size exponent for the polymer radius of gyration $R_{\mathrm{g}}$, we find that the upper boundary of the amplification regime ($R_{\mathrm{g}}$ exponent, $b \sim 3.6$) approaches exponents associated with dense packing, such as random Apollonian packings in three dimensions ($b \sim 3.5$) (see details for S5)~\cite{borkovec1994fractal,anishchik1995three}. This correspondence suggests that the enhanced overlap and entanglement concentrations in polymer solutions mirror the elevated jamming thresholds reported for athermal particles with power-law size distributions~\cite{shimamoto2023common,oquendo2020densest}.

Indeed, in our previous study of two-dimensional polydisperse particulate systems, power-law size distributions with a specific regime were shown to yield packing fractions exceeding the monodisperse limit, with maxima near exponents characteristic of Apollonian packing in two dimensions~\cite{shimamoto2023common}. The present results extend this concept to polymer systems, where entanglement and crowding replace direct geometric exclusion but lead to analogous collective constraints. From this perspective, the increase of $c^{\ast}$ and $c_{\mathrm{e}}$ under the specific $a$ regime can be interpreted as a shift of a jamming-like transition to higher concentrations, driven by continuous size polydispersity.

These findings point to a unifying physical principle: continuous, heavy-tailed polydispersity, encoded by a single exponent $a$, governs collective behaviors across disparate soft-matter systems. Whether in athermal particle packings or polymer solutions, the lower and upper boundaries of the specific $a$ regime determine how efficiently smaller components fill voids and how strongly larger components dominate collective constraints. This universality suggests that polydispersity is not a secondary complication but a fundamental organizing principle, particularly relevant to biological environments such as the cytoplasm, where macromolecular assemblies span broad, heavy-tailed size distributions. Recent studies indicate that molecular weight distribution regulates the equilibrium between biomolecular synthesis and degradation~\cite{chen2024viscosity}. Our results thus provide a quantitative bridge between polymer entanglement, soft packing, and jamming-like behavior in complex, highly polydisperse systems.

\section*{Conclusion}

We demonstrate that continuous molecular-weight polydispersity fundamentally reorganizes the collective behavior of polymer solutions. By systematically tuning the power-law exponent $a$, we identify a narrow regime ($1 < a \lesssim 2.5$) in which the viscosity scaling exponent in the entangled regime, the overlap concentration $c^{\ast}$, and the entanglement concentration ${c_{\mathrm{e}}}$ simultaneously exceed monodisperse limits. These enhanced responses arise from a balance between long polymers, which promote entanglement, and short polymers, which enable efficient void filling. Crucially, such behavior cannot be inferred from conventional descriptors such as the average molecular weight and the polydispersity index (PDI), but is encoded directly in the exponent $a$. More broadly, our results establish power-law polydispersity as a quantitative organizing principle for rheological behaviors in highly polydisperse systems, providing a physical link between entanglement, soft packing, and rheological amplification. Given that broad, heavy-tailed molecular-weight distributions naturally arise in cytoplasmic macromolecular assemblies, this framework offers a route toward understanding how intracellular environments tune their mechanics and biochemical functions through polydispersity. 

\section*{Supplementary Information}
Supplementary Materials provide additional experimental details, analyses, and supporting data. Section~S1 describes the materials used in this study, and Section~S2 details the gel permeation chromatography (GPC) measurements used to characterize molecular-weight distributions. Section~S3 explains the preparation protocol for constructing power-law molecular-weight distributions, while Section~S4 describes the absorbance measurements. Section~S5 presents the transformation from molecular-weight distributions to the corresponding radius-of-gyration distributions.

Figures~S1-S3 show the molecular-weight distributions of PEG samples and the mass fractions used to construct polydisperse ensembles with various power-law exponents. Figures~S4-S6 present additional rheological data, including shear-rate-dependent viscosity and concentration-dependent specific viscosity for monodisperse and polydisperse PEG samples with different upper molecular-weight cutoffs. Figures~S7 and~S8 examine the relationship between the power-law exponent $a$ and the fraction of long-chain polymers, as well as the dependence of the overlap and entanglement concentrations on $a$. Figure~S9 shows the wavelength and concentration dependence of absorbance for polydisperse PEG solutions.

\begin{acknowledgments}
This research was funded by the Japan Society for the Promotion of Science (JSPS) KAKENHI [grant nos. 23K22459 (M.Y.)], Japan Science and Technology Agency (JST) Program FOREST [JPMJFR213Y (M.Y.)], and Toyota Konpon Research Institute, Inc.
\end{acknowledgments}

\section*{AUTHORS CONTRIBUTIONS}
N.Y., D.S., and M.Y. conceived the project. 
N.Y. performed the experiments and conducted the formal analysis. 
N.Y. and M.Y. wrote the original draft. 
All authors contributed to the review and editing of the manuscript. 
M.Y. supervised the project and acquired funding.

\section*{COMPETING INTERESTS STATEMENT}
The authors declare that they have no competing interests. 

\section*{CORRESPONDENCE}
Correspondence and requests for materials should be addressed to N.~Y. (ynaoya@g.ecc.u-tokyo.ac.jp) and M.~Y (myanagisawa@g.ecc.u-tokyo.ac.jp).

\makeatletter
\renewcommand{\bibsection}{}
\makeatother
\section*{REFERENCES}
\bibliographystyle{naturemag}
\bibliography{poly-poly}

\clearpage

\end{document}


\title{Supplementary Information\\
\vspace{5mm}
Power-law molecular-weight distributions dictate universal behaviors in highly polydisperse polymer solutions}

\author{Naoya Yanagisawa$^*$}
\affiliation{Komaba Institute for Science, Graduate School of Arts and Sciences, The University of Tokyo, Komaba 3-8-1, Meguro, Tokyo 153-8902, Japan}

\author{Daisuke S. Shimamoto}
\affiliation{Komaba Institute for Science, Graduate School of Arts and Sciences, The University of Tokyo, Komaba 3-8-1, Meguro, Tokyo 153-8902, Japan}

\author{Miho Yanagisawa$^*$}
\affiliation{Komaba Institute for Science, Graduate School of Arts and Sciences, The University of Tokyo, Komaba 3-8-1, Meguro, Tokyo 153-8902, Japan}
\affiliation{Department of Physics, Graduate School of Science, The University of Tokyo, Hongo 7-3-1, Bunkyo, Tokyo 113-0033, Japan}
\affiliation{Center for Complex Systems Biology, Universal Biology Institute, The University of Tokyo, Komaba 3-8-1, Meguro, Tokyo 153-8902, Japan}

\date{\today}

\maketitle

\newpage
\tableofcontents

\newpage
\subsection*{\large S1: Materials}
Polydisperse polyethylene glycol (PEG) samples with molecular-weight distributions following power-law forms were prepared by mixing commercially available PEGs with different nominal molecular weights, listed below:
PEG 1k (Cat.\ No.\ 165-09085; FUJIFILM Wako Pure Chemical Co., Tokyo, Japan),
PEG 2k (Cat.\ No.\ 169-09105; FUJIFILM Wako Pure Chemical Co., Tokyo, Japan),
PEG 4.6k (Cat.\ No.\ 373001; Sigma-Aldrich, MO, USA),
PEG 6k (Cat.\ No.\ 169-09125; FUJIFILM Wako Pure Chemical Co., Tokyo, Japan),
PEG 8k (Cat.\ No.\ 194839; FUJIFILM Wako Pure Chemical Co., Tokyo, Japan),
PEG 12k (Cat.\ No.\ 81285; Sigma-Aldrich, MO, USA),
PEG 20k (Cat.\ No.\ 168-11285; FUJIFILM Wako Pure Chemical Co., Tokyo, Japan),
PEG 35k (Cat.\ No.\ 81310; Sigma-Aldrich, MO, USA),
PEO 100k (Cat.\ No.\ 181986; Sigma-Aldrich, MO, USA),
PEG 500k (Cat.\ No.\ 160-18521; FUJIFILM Wako Pure Chemical Co., Tokyo, Japan),
PEO 1000k (Cat.\ No.\ 043678.14; Thermo Fisher Scientific Inc., MA, USA),
and PEG 2000k (Cat.\ No.\ 166-13805; FUJIFILM Wako Pure Chemical Co., Tokyo, Japan).
As monodisperse reference samples, a PEO 786k GPC standard with a weight-average molecular weight $M_{\mathrm{w}} = 7.86 \times 10^{2}\,\mathrm{kg/mol}$ and a polydispersity index (PDI) of 1.16 was purchased from Tosoh Corporation (TSKgel standard PEO, Type SE-150).
Throughout this paper, polyethylene oxide (PEO) is denoted as PEG to avoid confusion.
PEG samples with PDI $\le 1.3$ are regarded as monodisperse.

\subsection*{\large S2: GPC Measurements}

\begin{figure*}[t]
\centering
\includegraphics[width=\textwidth]{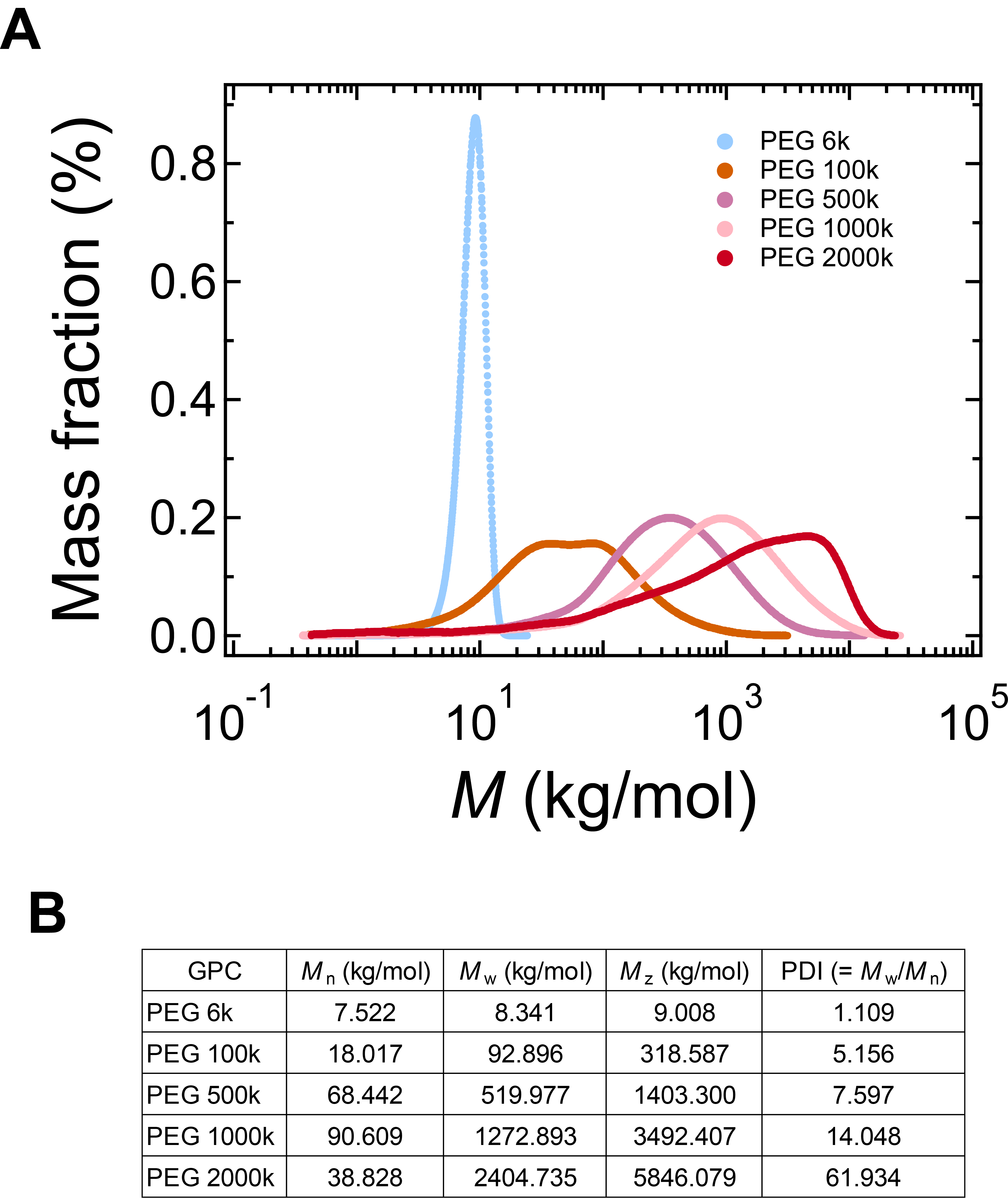}
\caption{(A) Weight-based molecular-weight distributions and (B) average molecular weight and polydispersity index (PDI) for PEG 6k, PEG 100k, PEG 500k, PEG 1000k, and PEG 2000k. $M_{\mathrm{z}}$ denotes the $z$-average molecular weight.
}
\label{FigS1}
\end{figure*}

Molecular-weight distributions were characterized by gel permeation chromatography (GPC) using an HLC-8320GPC system (Tosoh Co., Japan). Figure~\ref{FigS1} shows (A) the weight-based molecular-weight distributions and (B) the corresponding average molecular weights and polydispersity indices (PDI) for PEG 6k, PEG 100k, PEG 500k, PEG 1000k, and PEG 2000k. PEG 6k exhibits a narrow molecular-weight distribution with $\mathrm{PDI} \le 1.3$ and therefore corresponds to a monodisperse sample. In contrast, PEG samples with molecular weights of 100k and above show significantly broader molecular-weight distributions, indicating substantial polydispersity. Based on these observations, PEG samples with nominal molecular weights of 35k and below were assumed to follow Gaussian weight-based molecular-weight distributions with the reported molecular weight $M$ taken as the mean value. Specifically, the coefficient of variation $\sigma/\mu$ (where $\sigma$ is the standard deviation and $\mu$ is the mean) was assumed to be 0.28 for PEG 1k ($\mathrm{PDI}=1.1$), 0.30 for PEG 2k ($\mathrm{PDI}=1.1$), 0.26 for PEG 4.6k ($\mathrm{PDI}=1.1$), 0.25 for PEG 8k ($\mathrm{PDI}=1.1$), 0.23 for PEG 12k ($\mu = 13\ \mathrm{kg/mol}$, $\mathrm{PDI}=1.27$), 0.24 for PEG 20k ($\mathrm{PDI}=1.3$), and 0.24 for PEG 35k ($\mathrm{PDI}=1.3$). The validity of these assumptions is supported by the agreement between the calculated molecular-weight distributions for $a = 1$ shown as black symbols and the experimentally measured GPC data shown as red symbols in Fig.~1C of the main text.

\subsection*{\large S3: Preparation of power-law molecular-weight distributions}

Power-law molecular-weight distributions were prepared by blending the commercially available PEG samples ranging from PEG 1k to PEG 2000k at prescribed mass fractions (see Fig.~S2). The resulting molecular-weight distributions follow a power-law form with exponent $a$, as shown in Fig.~1C and Fig.~S3. Throughout this study, the minimum molecular weight was fixed at $M_{\mathrm{min}}$ = 1 kg/mol. Power-law distributions were constructed for several upper cutoffs of the molecular weight, namely $M_{\mathrm{max}} \sim 1 \times 10^{4}$, $1 \times 10^{3}$, $2 \times 10^{2}$, and $45\ \mathrm{kg/mol}$. Except for the case of $M_{\mathrm{max}} \sim 45\ \mathrm{kg/mol}$, the exponent $a$ was systematically varied in the range $0.75 \le a \le 2.5$. Values of $a < 0.75$ could not be realized experimentally because PEG samples with molecular weights of 100k and above contain a substantial fraction of short-chain components, which prevents the formation of sufficiently shallow distributions. Conversely, values of $a > 2.5$ were experimentally inaccessible because the contribution of minor components becomes vanishingly small.

\subsection*{\large S4: Absorbance Measurements}

To examine the presence or absence of spatial heterogeneity in structures in PEG solutions, wavelength-dependent absorbance measurements were carried out at $25\,^{\circ}\mathrm{C}$ using a UV-visible spectrophotometer (UV-1900i, Shimadzu Co., Japan). Quartz cuvettes with an optical path length of 10~mm (10~$\times$~10~$\times$~40~mm) were used.
Figure~S9 shows the concentration dependence of the absorbance for samples with $M_{\mathrm{max}} \sim 1 \times 10^{3}\ \mathrm{kg/mol}$ at $a = 1.5$, which corresponds to the peak region of the scaling exponent of the specific viscosity above the entanglement concentration. The dashed line indicates the entanglement concentration $c_{\mathrm{e}}$ determined from rheological measurements. The concentration at which the absorbance exhibits a pronounced increase coincides with $c_{\mathrm{e}}$.
The increase in absorbance observed above the entanglement concentration suggests the formation of heterogeneity in polymer density induced by polydispersity, originating from enhanced light scattering.

\subsection*{\large S5: Transformation from Molecular-Weight Distributions to Radius-of-Gyration Distributions}

Here, in order to compare with random packing of three-dimensional spheres, we transform the molecular-weight distribution $N(M) \propto M^{-a}$ into a distribution of the radius of gyration $R_{\mathrm{g}}$.
For PEG in dilute solution, the radius of gyration is known to scale with molecular weight as
\begin{equation}
R_{\mathrm{g}} = 0.0215\, (1000M)^{0.583}\ \mathrm{nm},
\end{equation}
where $M$ is given in units of kg/mol~\cite{wisniewska2014scaling,devanand1991asymptotic}.
The molecular-weight distribution $N(M)$ can be transformed into the distribution of the radius of gyration $N(R_{\mathrm{g}})$ through a change of variables according to
\begin{equation}
N(R_{\mathrm{g}}) = N(M)\left|\frac{dM}{dR_{\mathrm{g}}}\right|.
\end{equation}
Assuming a scaling relation between the radius of gyration and molecular weight,
\begin{equation}
R_{\mathrm{g}} \propto M^{\nu},
\end{equation}
with $\nu = 0.583 \simeq 1/1.715$, the molecular weight can be expressed as $M \propto R_{\mathrm{g}}^{1.715}$.
For a power-law molecular-weight distribution $N(M) \propto M^{-a}$, this transformation yields
\begin{align}
N(R_{\mathrm{g}})
&\propto \left(R_{\mathrm{g}}^{1.715}\right)^{-a}
\frac{d}{dR_{\mathrm{g}}}\left(R_{\mathrm{g}}^{1.715}\right) \\
&\propto R_{\mathrm{g}}^{-1.715a} R_{\mathrm{g}}^{0.715} \\
&\propto R_{\mathrm{g}}^{-(1.715a - 0.715)} \\
&\propto R_{\mathrm{g}}^{-b},
\end{align}
where the exponent $b$ is given by
\begin{equation}
b = 1.715a - 0.715.
\end{equation}
For $1 < a \leq 2.5$ and the peak at $a=1.5$ with $M_{\mathrm{min}} = 10^{0}$ and $M_{\mathrm{max}} = 10^{4}$~kg/mol, this corresponds to the range $1 < b \leq 3.6$ with the peak at $b \simeq 1.86$.
The corresponding bounds of the radius of gyration are $R_{\mathrm{g,min}} \simeq 1.2$~nm and $R_{\mathrm{g,max}} \simeq 259$~nm.
Notably, the upper bound of the exponent $b \simeq 3.6$ is close to the Apollonian packing exponent reported for three-dimensional spheres ($b \simeq 3.5$)~\cite{borkovec1994fractal,anishchik1995three}.

\vspace{50mm}

\subsection*{\large Supplementary Figures}

\clearpage

\begin{figure*}[t]
\centering
\includegraphics[width=\textwidth]{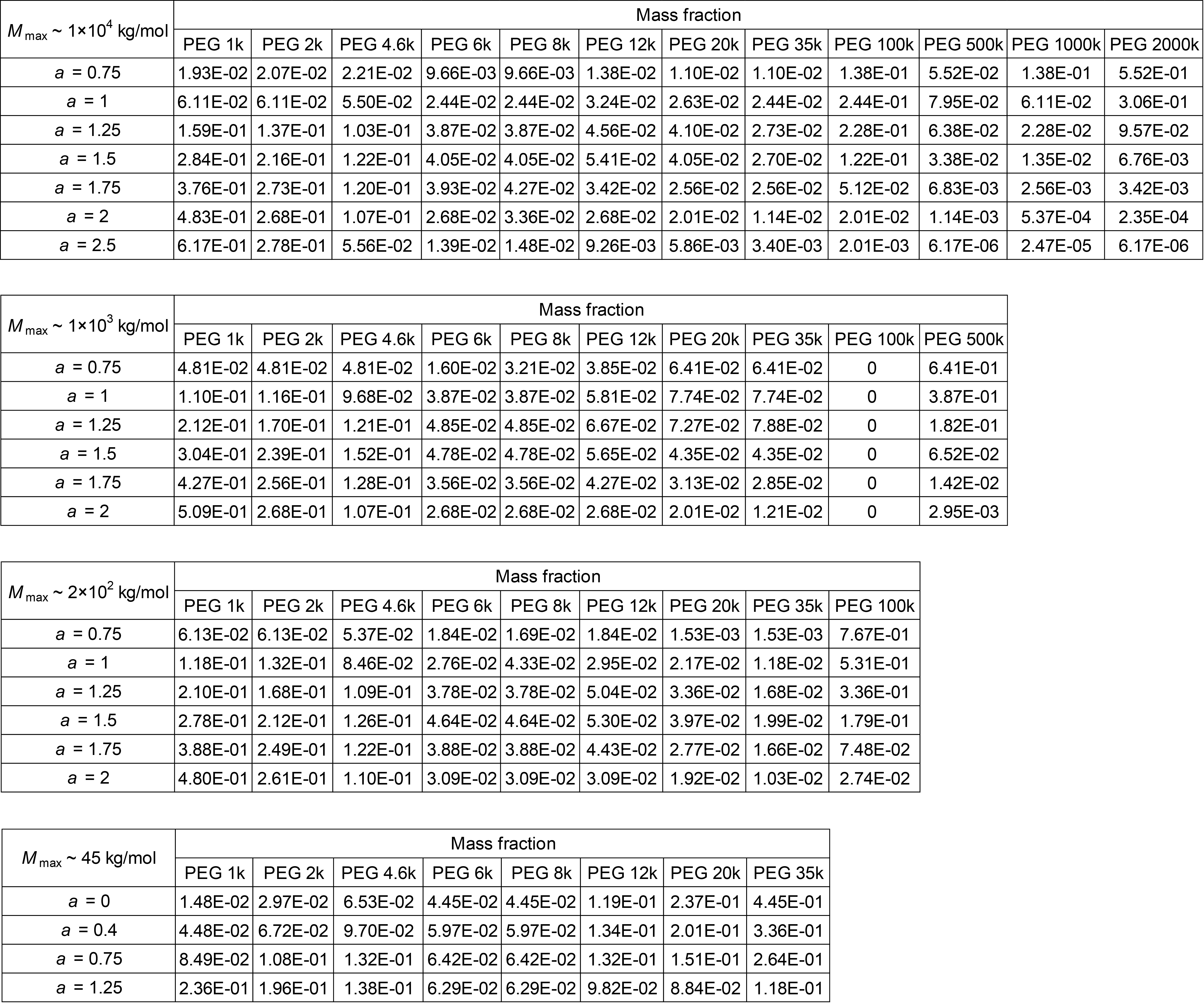}
\caption{Mass fractions of individual PEG components used to construct power-law molecular-weight distributions with different exponents $a$ and upper molecular-weight cutoffs $M_{\mathrm{max}}$.
}
\label{FigS2}
\end{figure*}

\begin{figure*}[t]
\centering
\includegraphics[width=\linewidth]{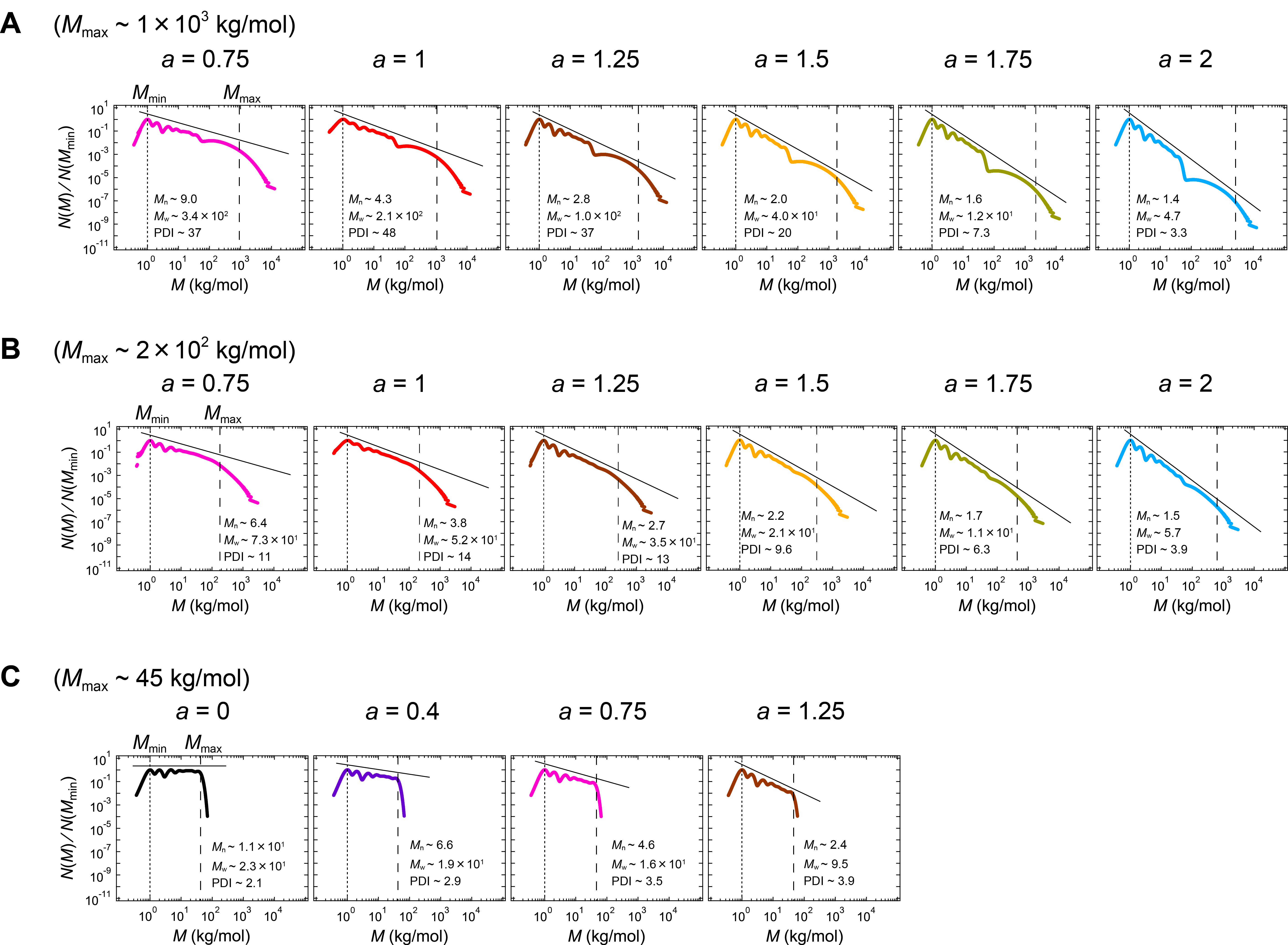}
\caption{(A-C) Power-law molecular-weight distributions with different exponents $a$ for upper molecular-weight cutoffs of $M_{\mathrm{max}} \sim 1 \times 10^{3}$, $2 \times 10^{2}$, and $45\ \mathrm{kg/mol}$, respectively.
}
\label{FigS3}
\end{figure*}

\begin{figure*}[t]
\centering
\includegraphics[width=90mm]{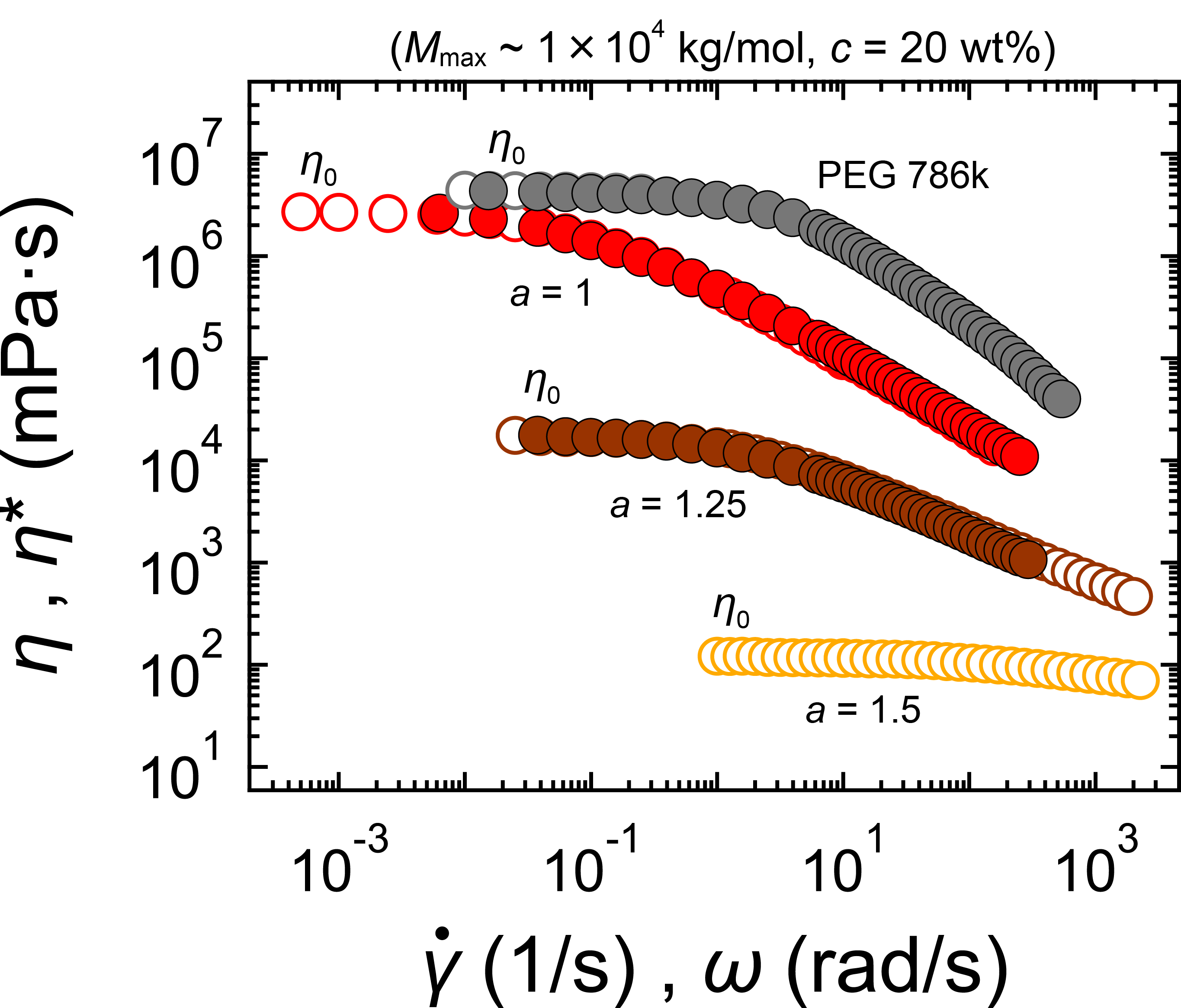}
\caption{Representative flow curves showing the shear-rate-dependent viscosity $\eta$ (open symbols) and the angular-frequency-dependent complex viscosity $\eta^{*}$ (filled symbols).
}
\label{FigS4}
\end{figure*}

\begin{figure*}[t]
\centering
\includegraphics[width=90mm]{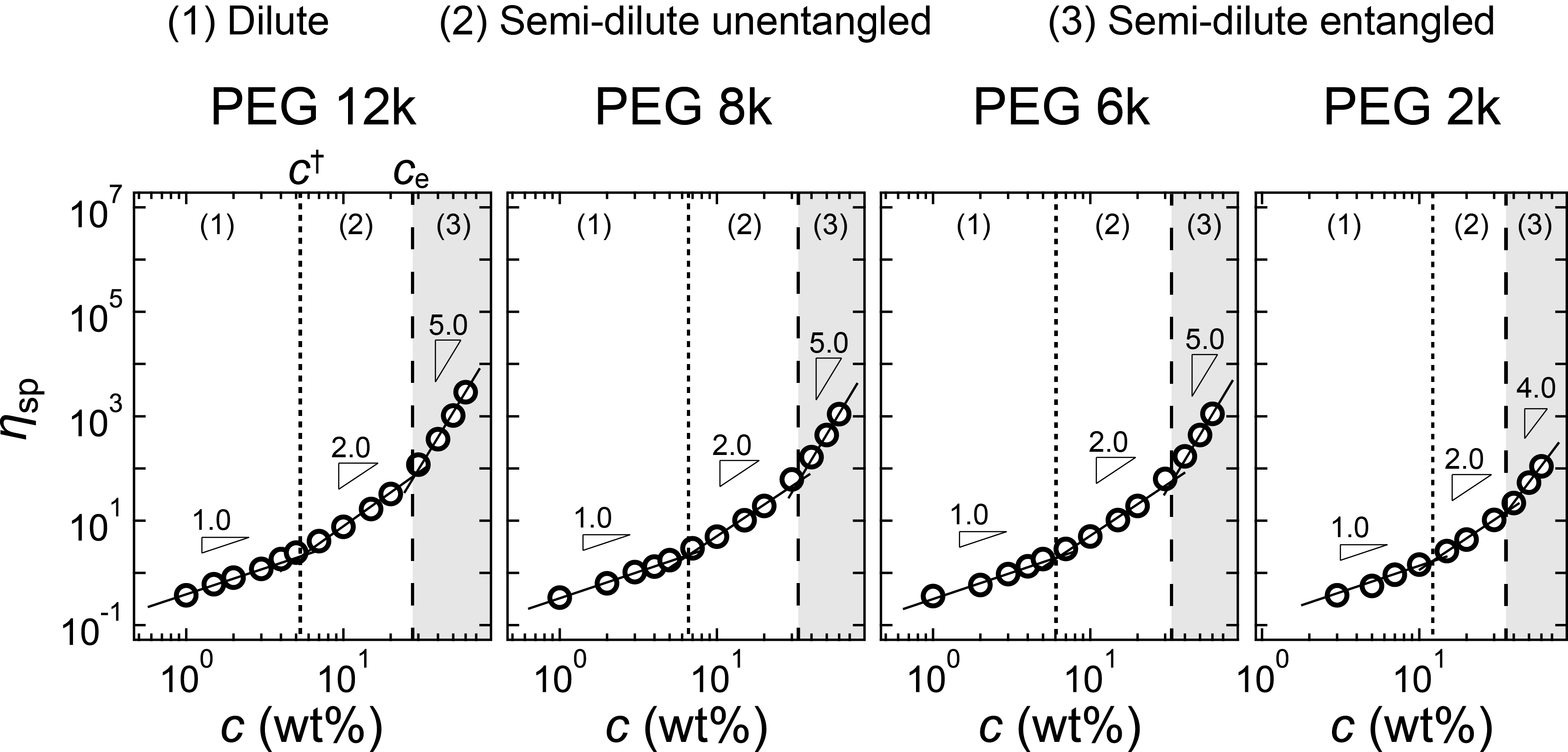}
\caption{Concentration dependence of the specific viscosity $\eta_{\mathrm{sp}}$ for monodisperse samples.
}
\label{FigS5}
\end{figure*}

\begin{figure*}[t]
\centering
\includegraphics[width=\linewidth]{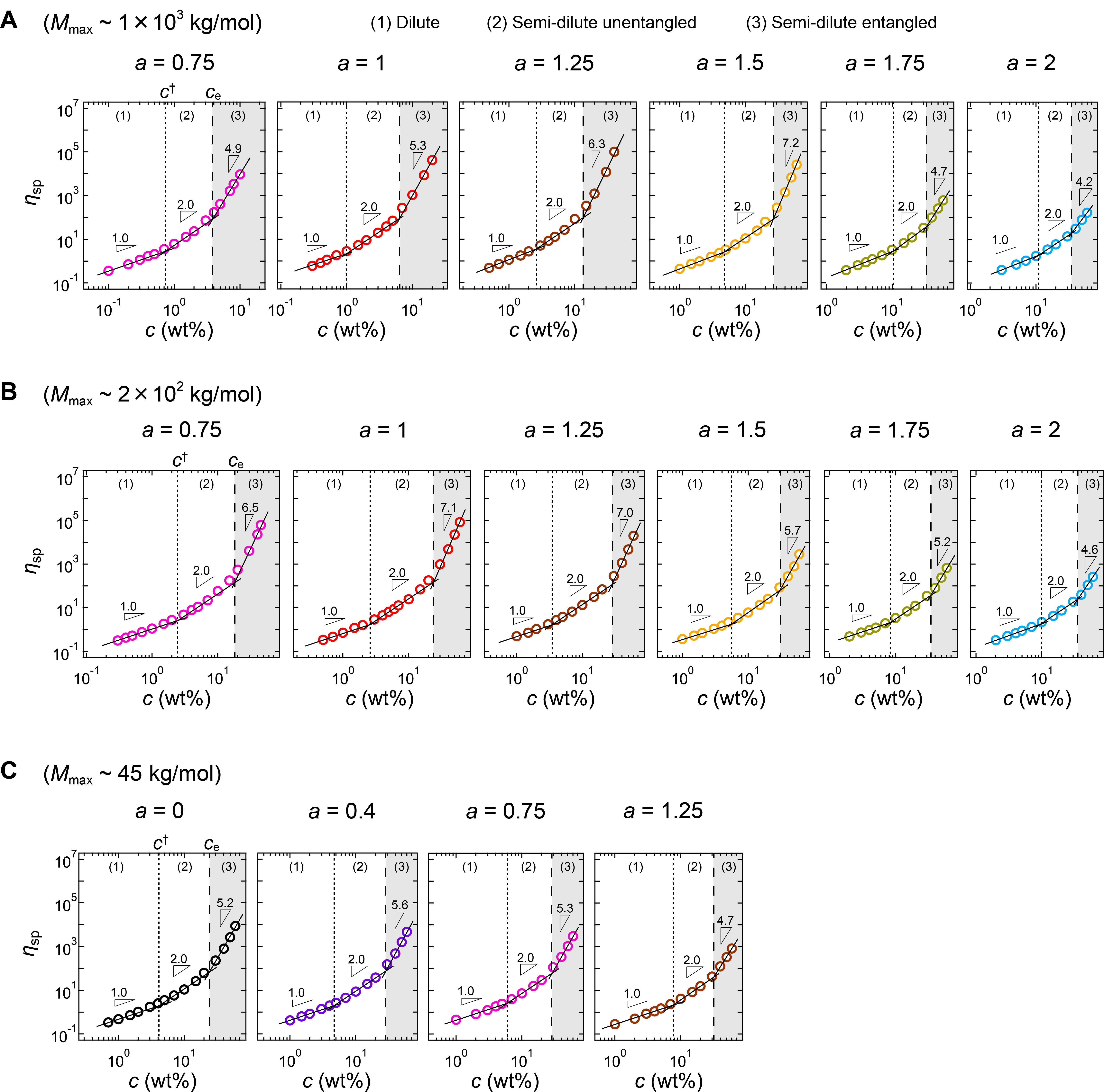}
\caption{(A-C) Concentration dependence of the specific viscosity $\eta_{\mathrm{sp}}$ for polydisperse samples with upper molecular-weight cutoffs of $M_{\mathrm{max}} \sim 1 \times 10^{3}$, $2 \times 10^{2}$, and $45\ \mathrm{kg/mol}$, respectively.
}
\label{FigS6}
\end{figure*}

\begin{figure*}[t]
\centering
\includegraphics[width=\linewidth]{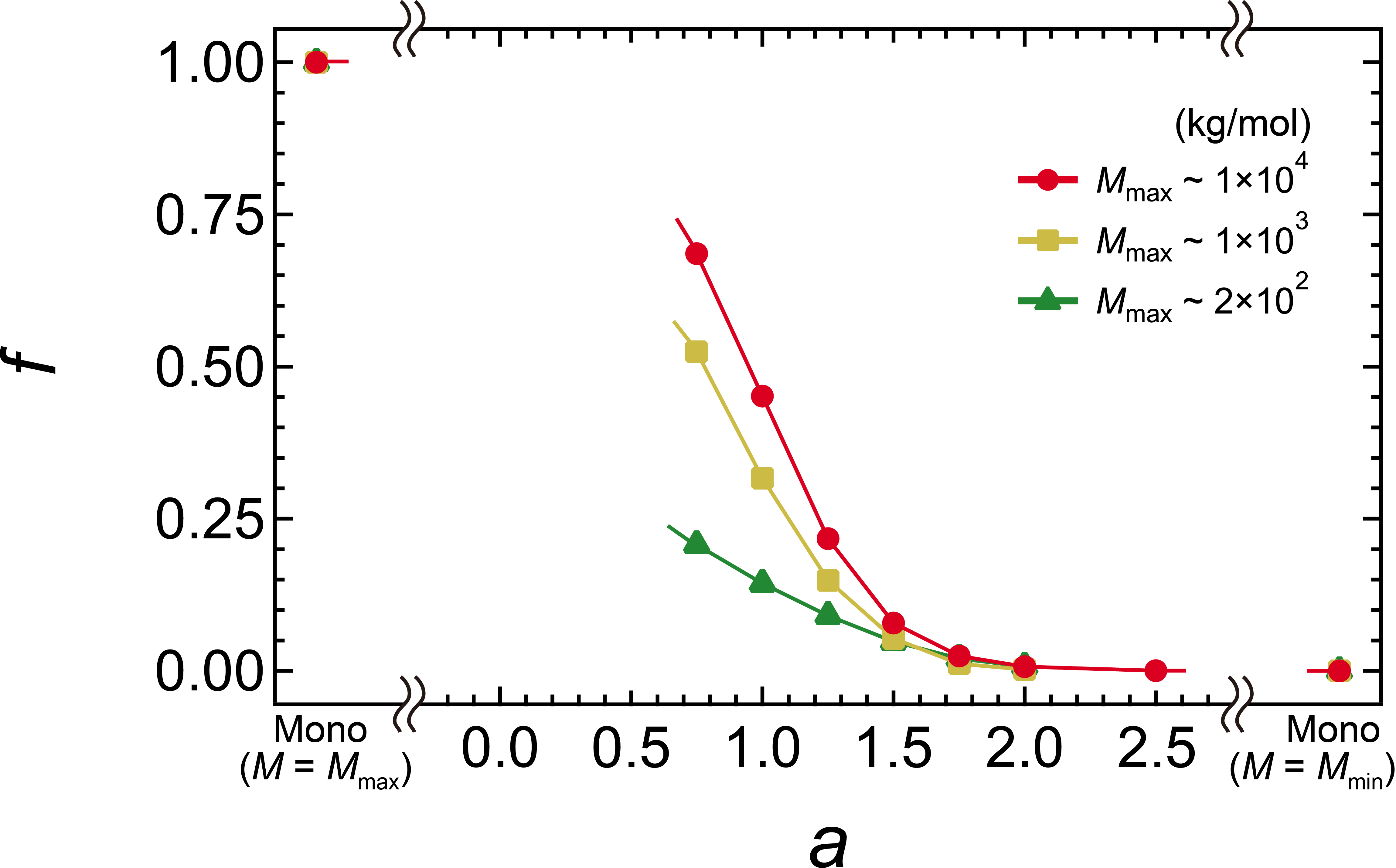}
\caption{Relationship between the mass proportion of long-polymer components ($M > 100$~kg/mol) capable of forming entanglements $f$ and $a$.
}
\label{FigS7}
\end{figure*}

\begin{figure*}[t]
\centering
\includegraphics[width=\linewidth]{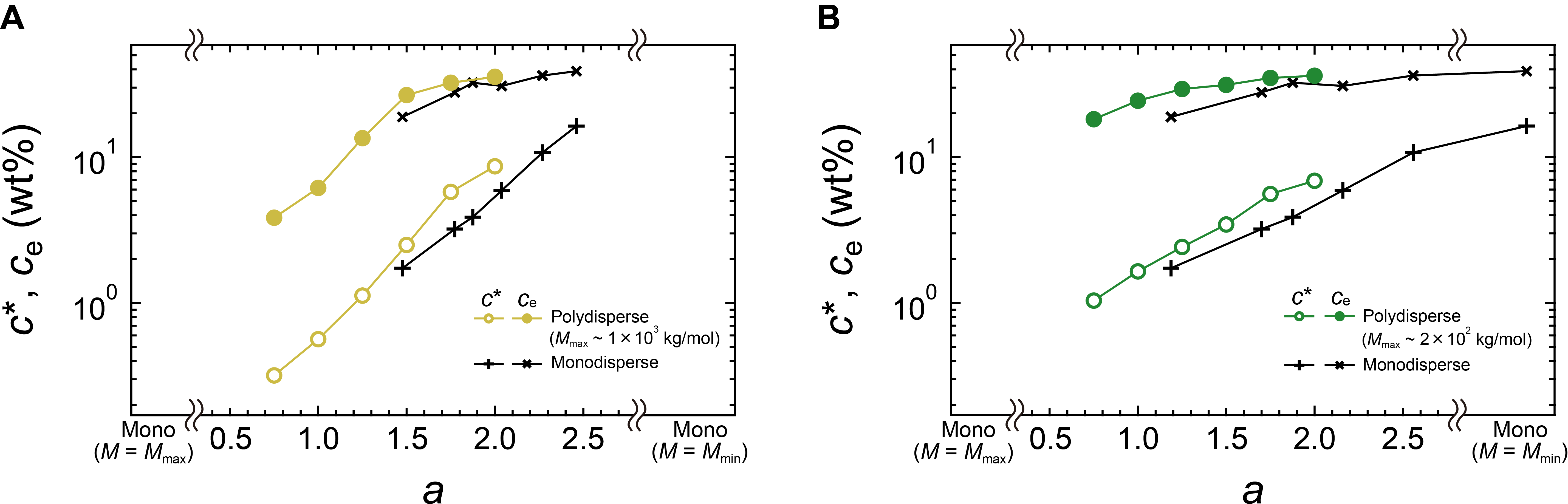}
\caption{Dependence of the overlap concentration $c^{\ast}$ and the entanglement concentration $c_{\mathrm{e}}$ on the exponent $a$ for (A) $M_{\mathrm{max}} \sim 1 \times 10^{3}$ and (B) $2 \times 10^{2}\ \mathrm{kg/mol}$. Both $c^{\ast}$ and $c_{\mathrm{e}}$ exceed the corresponding values for monodisperse samples in the range of $a$ where the scaling exponent of the specific viscosity above $c_{\mathrm{e}}$ is enhanced.
}
\label{FigS8}
\end{figure*}

\begin{figure*}[t]
\centering
\includegraphics[width=\linewidth]{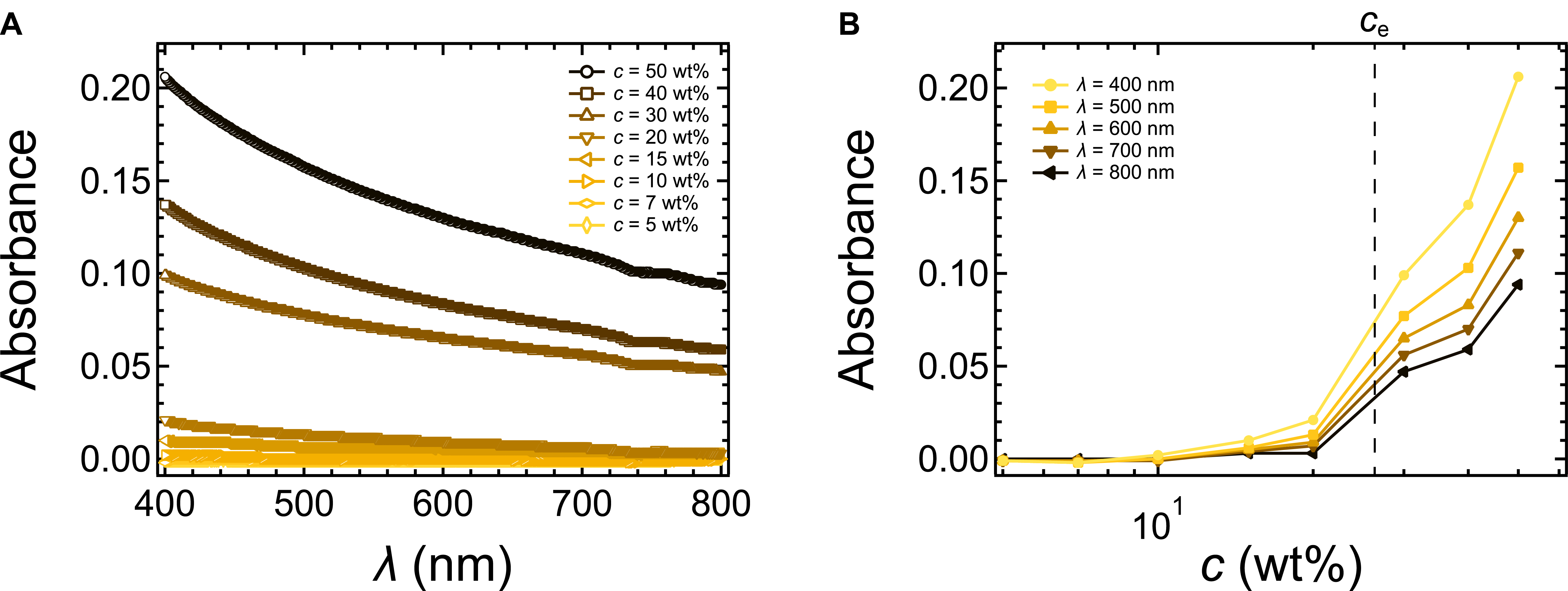}
\caption{(A) Wavelength dependence and (B) concentration dependence of the absorbance for a polydisperse PEG solution with $a = 1.5$ and $M_{\mathrm{max}} \sim 1 \times 10^{3}\ \mathrm{kg/mol}$. The dashed line in (B) indicates the entanglement concentration $c_{\mathrm{e}}$ obtained from rheological measurements, which coincides with the concentration at which the absorbance begins to increase sharply.
}
\label{FigS9}
\end{figure*}

\clearpage

\makeatletter
\renewcommand{\bibsection}{}
\makeatother
\subsection*{\large References}
\bibliographystyle{naturemag}
\bibliography{poly-poly}

\clearpage